\documentclass[twocolumn,prl,superscriptaddress,showpacs]{revtex4}
\usepackage{amsmath}
\usepackage{graphicx}

\newcommand{\ket}[1]{\ensuremath{|#1\rangle}}
\newcommand{\bra}[1]{\ensuremath{\langle#1|}}
\newcommand{\proj}[1]{\ket{#1}\bra{#1}}
\newcommand{\nuc}[2]{\ensuremath{{}^{#1}{\text{#2}}}}
\newcommand{\Tone}{\ensuremath{\text{T}_1}}
\newcommand{\Ttwo}{\ensuremath{\text{T}_2}}
\newcommand{\DOD}{\ensuremath{\text{D}_2\text{O}}}
\newcommand{\CNOT}{controlled-\textsc{not}}
\newcommand{\UNOT}{universal-\textsc{not}}

\begin{document}
\title{Approximate Quantum Cloning with Nuclear Magnetic Resonance}
\author{Holly K. Cummins}
\affiliation{Centre for Quantum Computation, Clarendon Laboratory,
University of Oxford, Parks Road, OX1 3PU, United Kingdom}
\author{Claire Jones}
\affiliation{Dyson Perrins Laboratory, University of Oxford, South
Parks Road, OX1 3QY, United Kingdom}
\author{Alistair Furze}
\affiliation{Dyson Perrins Laboratory, University of Oxford, South
Parks Road, OX1 3QY, United Kingdom}
\author{Nicholas F. Soffe}
\affiliation{Oxford Centre for Molecular Sciences, Central
Chemistry Laboratory, University of Oxford, South Parks Road, OX1
3QTH, United Kingdom}
\author{Michele Mosca}
\affiliation{Department of Combinatorics and Optimization,
University of Waterloo, Waterloo, Ontario N2L 3G1, Canada}
\author{Josephine M. Peach}
\affiliation{Dyson Perrins Laboratory, University of Oxford, South
Parks Road, OX1 3QY, United Kingdom}
\author{Jonathan A. Jones}\email{jonathan.jones@qubit.org}
\affiliation{Centre for Quantum Computation, Clarendon Laboratory,
University of Oxford, Parks Road, OX1 3PU, United
Kingdom}
\affiliation{Oxford Centre for Molecular Sciences, Central
Chemistry Laboratory, University of Oxford, South Parks Road, OX1
3QTH, United Kingdom}
\date{\today}
\begin{abstract}
Here we describe a Nuclear Magnetic Resonance (NMR) experiment
that uses a three qubit NMR device to implement the one to two
approximate quantum cloning network of Bu\v{z}ek \textit{et al}.
\end{abstract}
\maketitle Quantum information processing \cite{bennett00} has
been the subject of much recent interest, not only because it
offers new modes of computation and communication, but also
because quantum information differs from classical information in
several fundamental ways. One important example is the fact that
it is impossible to accurately clone (copy) an unknown quantum
state \cite{wooters82}, and so quantum bits (qubits) cannot be
duplicated.  It is, however, possible to prepare an approximate
copy \cite{buzek96}, and several schemes for optimal approximate
cloning have been developed. Nuclear Magnetic Resonance (NMR)
\cite{ernst, hore, claridge} has already been used to demonstrate
simple quantum information processing methods \cite{cory96,
cory97, gershenfeld97, jones98, chuang98, chuang98b, jones98b,
linden98, nielsen98, cory98, somaroo99, jones99, dorai99, marx00,
vandersypen00, knill00, knill01, jones01}, and here we describe an
NMR experiment that uses a three qubit NMR device to implement the
one to two approximate cloning network \cite{buzek97} of Bu\v{z}ek
\textit{et al}.

\begin{figure}
\includegraphics{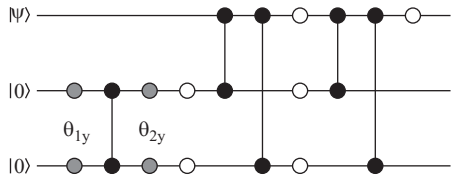}
\caption{\label{fig:1}Our slightly modified version of the
approximate quantum cloning network developed by Bu\v{z}ek
\textit{et al}.  The initial ``preparation'' stage has been
replaced by an alternative network, which is simpler to implement
with NMR techniques; the second ``copy'' stage is unchanged.
Filled circles connected by control lines indicate controlled
$\pi$ phase shift gates \cite{jones01}, empty circles indicate
single qubit Hadamard gates, while grey circles indicate other
single qubit rotations. The two rotation angles in the preparation
stage are $\theta_1=\arcsin\left(1/\sqrt{3}\right)\approx
35^\circ$ and $\theta_2=\pi/12=15^\circ$.}
\end{figure}

We started by slightly modifying the approximate cloning network
\cite{buzek97} of Bu\v{z}ek \textit{et al.} to take advantage of
our specific hardware; our version of the network is shown in
Fig.~\ref{fig:1}.
This takes in one qubit in an arbitrary state (normally considered
to be a pure state, \ket{\psi}), and two ancilla qubits in state
\ket{0}; the two ancilla qubits are prepared into an appropriate
initial state, and then the input qubit is copied.  At the end of
the cloning sequence the three qubits are all entangled with one
another, and so the reduced density operator descriptions of each
qubit (obtained by tracing out the other two qubits) correspond to
mixed states.  The two ancilla qubits are both in the state
$\frac{5}{6}\proj{\psi}+\frac{1}{6}\proj{\psi^\perp}$ (that is,
approximate clones of \ket{\psi} with fidelity $5/6$), while the
input qubit is now the approximate transpose of \proj{\psi}.
Alternatively, using the identity
$\proj{\psi}+\proj{\psi^\perp}=\mathbf{1}$, the states of the
ancilla qubits can be written as
$\frac{2}{3}\proj{\psi}+\frac{1}{3}\mathbf{1}/2$. Cloning is not,
of course, confined to pure states, and any mixed state $\rho$ can
be cloned to give $\frac{2}{3}\rho+\frac{1}{3}\mathbf{1}/2$ . Such
mixed state cloning is particularly easy to study using NMR
techniques, as the ensemble averaging inherent in NMR experiments
is a help rather than a hindrance.

\begin{figure}
\includegraphics[scale=0.4]{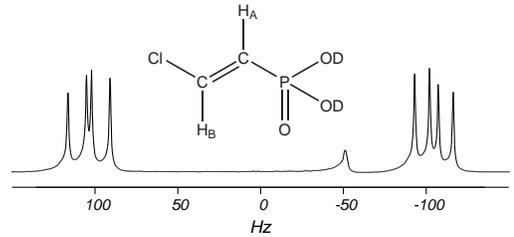}
\caption{\label{fig:2}The three qubit system provided by
E-(2-chloro\-ethenyl)\-phos\-phonic acid dissolved in
$\text{D}_2\text{O}$ and its \nuc{1}{H} NMR spectrum.  The broad
peak near $-50$ Hz is a folded signal arising from residual HOD.}
\end{figure}

We implemented this network using a 3 qubit NMR quantum computer
based on the single \nuc{31}{P} nucleus ($P$) and the two
\nuc{1}{H} nuclei ($A$ and $B$) in
E-(2-chloro\-ethenyl)\-phos\-phonic acid (Fig.~\ref{fig:2})
dissolved in \DOD.  Ethynylphosphonic acid diethyl ester was
prepared from trimethylsilylacetylene by successive treatment with
EtMgBr and diethyl chlorophosphate, followed by cold aqueous
$\text{Na}_2\text{CO}_3$ to remove the silyl protecting group
\cite{monaghan83}. HCl was added \textit{cis} across the triple
bond using LiCl in acetic acid \cite{huang95}, and the Z--E
isomerisation was accomplished by heating with PhSH in the
presence of AIBN \cite{danieski96}. Acidic hydrolysis of the ester
groups gave E-(2-chloro\-ethenyl)\-phos\-phonic acid (overall
yield 5\%, purity $>95$\%), from which contaminating metal ions
were removed using chelex resin. The NMR sample was prepared by
dissolving 7mg of compound into 600$\mu$L of \DOD\ to give a 0.08M
solution, and all NMR experiments were run at a nominal
temperature of $20^\circ$ using a homebuilt 600 MHz (\nuc{1}{H}
frequency) NMR spectrometer at the OCMS with a homebuilt double
resonance (\nuc{1}{H} inner, \nuc{31}{P} outer) probe.  The
measured NMR parameters are listed in Table~\ref{table:1}.
\begin{table}
\caption{\label{table:1}Measured NMR parameters for
E-(2-chloro\-ethenyl)\-phos\-phonic acid dissolved in \DOD\
(0.08M, $20^\circ$). Frequencies are measured with respect to
transmitter frequencies of 600.1517482 MHz (\nuc{1}{H}) and
242.9458642 MHz (\nuc{31}{P}). {\Tone} relaxation times were
measured by inversion-recovery methods, while {\Ttwo} times were
measured using a single spin-echo; all relaxation times are
averaged over the four components of the relevant multiplet, as
there was little sign of correlated relaxation.}
\begin{ruledtabular}
\begin{tabular}{lrrrrrr}
&$\nu$/Hz&{\Tone}/s&{\Ttwo}/s&$J_P$/Hz&$J_A$/Hz&$J_B$/Hz\\
$P$&0.0&3.8&0.72&&9.1&11.3\\
$A$&104.0&17.6&1.82&9.1&&14.3\\
$B$&$-104.0$&16.9&1.82&11.3&14.3&
\end{tabular}
\end{ruledtabular}
\end{table}

\begin{figure}
\includegraphics[scale=0.95]{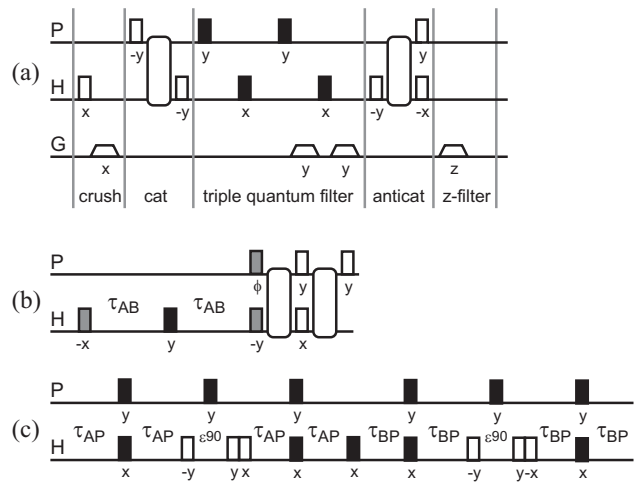}
\caption{\label{fig:3}NMR pulse sequences. White and black boxes
are $90^\circ$ and $180^\circ$ pulses, while grey boxes are pulses
with other flip angles; pulse phases and gradient directions are
shown below each pulse.  The large ovals correspond to the
``echo'' sequence described below.  All RF pulses are hard, with
\nuc{1}{H} frequency selection achieved using ``jump and return''
sequences \cite{jones99, plateau82}. The purification sequence (a)
and cloning sequence (b) are built around the echo sequence (c)
which implements the coupling element of $PA$ and $PB$ controlled
phase shifts \cite{jones01}.  Delay times are
$\tau_{AB}=1/(4J_{AB})$, $\tau_{AP}=1/(8J_{AP})$ and
$\tau_{BP}=1/(8J_{BP})$; $\epsilon_{90}$ was set as described in
the main text.}
\end{figure}

This system may be conveniently described using product operator
notation \cite{sorensen83}. The experiment begins with the
preparation of the initial state $P_zA_0B_0$ using a modification
of the ``cat state'' method of Knill \textit{et al.}
\cite{knill00}. Qubit $P$ can then be set to any desired point on
the Bloch sphere using a single radiofrequency (RF) pulse.  After
the cloning sequence, the reduced density matrices of qubits $A$
and $B$ will correspond to Bloch vectors lying parallel to the
original Bloch vector of $P$, but with lengths only $2/3$ that of
the original vector; note that the effect of the maximally mixed
component is simply to reduce the length of the Bloch vector
without affecting its orientation. Detailed pulse sequences are
shown in Fig.~\ref{fig:3} and described below.

NMR pulse sequences were developed by replacing the abstract gates
in Fig.~\ref{fig:1} with an idealized sequence of NMR pulses
(including $z$-rotations) and delays; the resulting sequences were
then simplified by adsorbing $z$-rotations into abstract reference
frames \cite{knill00, jones01} and combining RF pulses when
convenient \cite{jones01}. The detailed implementation of the
sequence was chosen so as to avoid \nuc{1}{H} selective pulses as
far as possible; those selective pulses in the ``echo'' sequence
that could not be avoided were implemented using a variant on the
``jump and return'' sequence \cite{jones99, plateau82}. This
sequence incorporates a delay which should na\"{\i}vely take the
value $\epsilon_{90}=1/(4\delta\nu)$, where the two \nuc{1}{H}
signals have frequencies $\pm\delta\nu$, but in practice it is
better to slightly reduce its length in order to allow for
non-idealities arising from the fact that $\delta\nu$ is not much
greater than the $J$ couplings; in our experiments $\epsilon_{90}$
was reduced to 90\% of its nominal value.

The initial ``purification'' sequence was developed separately,
but uses many of the same ideas.  Our implementation is based on
the ``cat state'' methods \cite{knill00} of Knill \textit{et al}.
The core of the sequence comprises a Hadamard gate and a pair of
\CNOT\ gates that convert $P_z$ to a density matrix containing all
terms along the anti-diagonal, a gradient filter that selects the
desired triple quantum term, and a second pair of \CNOT\ gates and
a Hadamard which act to produce $P_zA_0B_0$. The novel gradient
filter sequence used here uses two gradients with strengths in the
ratio $1:0.6633$ to select the desired three quantum terms while
crushing other anti-diagonal terms; note that phase cycling is
\textit{not} required.

Unlike conventional gradient sequences used to select coherence
transfer pathways \cite{hore, knill00}, this scheme is largely
unaffected by diffusion. It is, however, vulnerable to
imperfections in the anticat sequence: these are removed by the
final $z$-filter, which incorporates a variable delay to suppress
zero quantum terms \cite{jones99}.  Note that this variable delay
can be combined with the standard \textsc{cyclops} phase cycle
\cite{claridge}, and so no additional phase cycling is necessary
\cite{jones99}. In principle this sequence will also suppress
terms arising from initial $A_z$ and $B_z$ magnetization, as well
as signal from residual HOD and other impurities; in reality the
behaviour is not as selective as one might desire, and so the
purification sequence is preceded by a $90^\circ$ \nuc{1}{H}
excitation pulse and gradient crush, to reduce any signals that do
not start from $P_z$.

The effects of the idealized network (Fig.~\ref{fig:1}) on an NMR
system with spin $P$ set to an arbitrary point on the Bloch
sphere, $P_{\theta\phi}$, are easily calculated: the observable
NMR signal takes the form
\begin{equation}
\frac{1}{3}\left[A_{\theta\phi} \left( B_0P_0+B_1P_1\right) +
B_{\theta\phi} \left(A_0P_0+A_1P_1\right) \right].
\end{equation}
After tracing out the other two spins the signal from spins $A$
and $B$ is reduced to the expected forms,
$\frac{2}{3}A_{\theta\phi}$ and $\frac{2}{3}B_{\theta\phi}$
respectively.  Note that the component corresponding to the
maximally mixed state has no effect on the NMR signal as all NMR
observables are traceless. Spin $P$ could be traced out
experimentally by applying a \nuc{31}{P} decoupling field during
observation of the two \nuc{1}{H} nuclei, $A$ and $B$, but this
approach was found to lead to unacceptable sample heating, with
resultant shifts in resonance frequencies. Similarly it would in
principle be possible to trace out spin $A$ while observing $B$ by
the use of selective homonuclear decoupling, but this has the
further disadvantage of requiring two separate experiments to
observe spins $A$ and $B$. For these reasons we decided to
implement the tracing out process in software by adding together
signals arising from the four components in each multiplet; this
is most simply achieved by integrating the entire multiplet.

\begin{figure}
\includegraphics[scale=0.4]{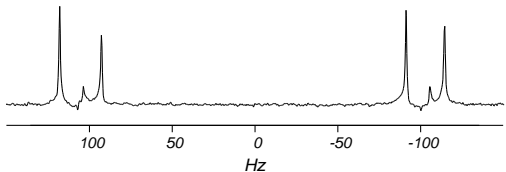}
\caption{\label{fig:4}The experimental result from cloning the
initial state $P_x$.  The receiver phase was set using a separate
experiment so that $x$-magnetization appears as positive
absorption mode lines.  The NMR experiment was repeated four times
using the \textsc{cyclops} phase cycle \cite{claridge} to reduce
instrumental imperfections.}
\end{figure}

The experimental results from cloning the initial state $P_x$ are
shown in Fig.~\ref{fig:4}.
The spectrum does have the expected overall form (inphase
absorption signals at the outermost positions of each multiplet),
but some non-idealities are clearly visible: the two lines in each
multiplet have different heights, and there are small signals at
other positions in the multiplets. More seriously the observed
signal intensity is only about 25\% of the expected value.
Detailed calculations (data not shown) indicate that the errors in
relative line intensities can be traced to $J$ coupling evolution
during selective excitation sequences, while the overall intensity
loss can be ascribed to \Ttwo\ relaxation and $\text{B}_1$
inhomogeneity effects.

\begin{figure*}
\includegraphics[scale=0.667]{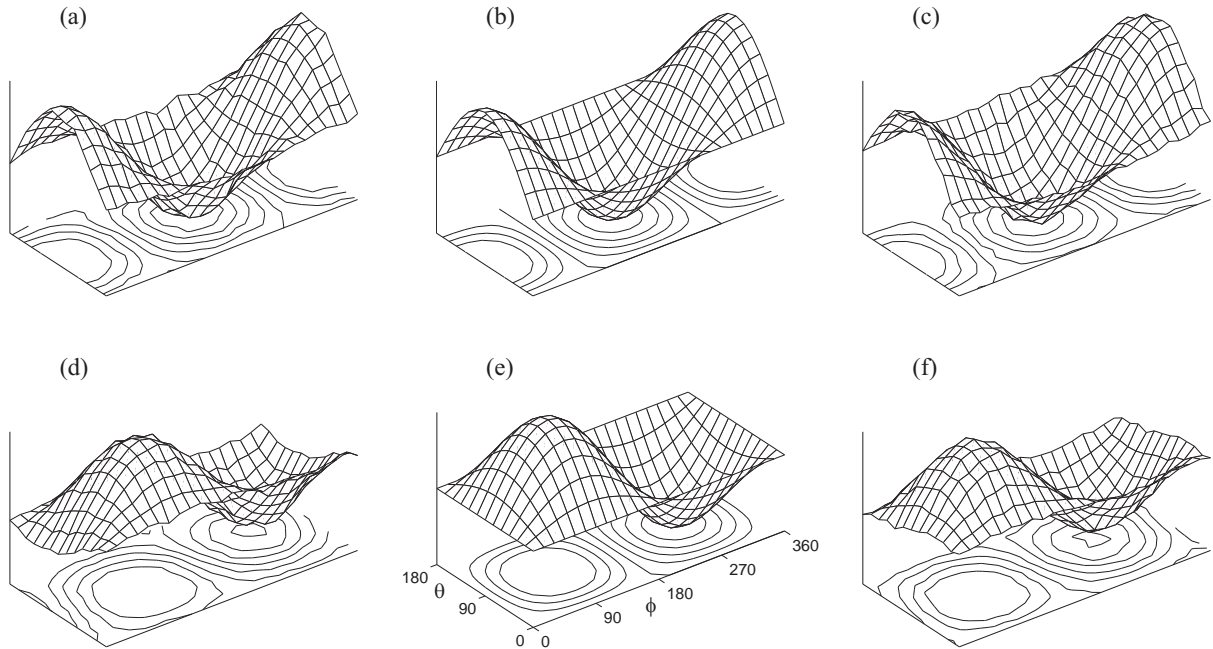}
\caption{\label{fig:5}Summarised experimental results from cloning
the general initial state $P_x \sin\theta\cos\phi +
P_y\sin\theta\sin\phi + P_z\cos\theta$; the integrals of the real
(a, b, c) and imaginary (d, e, f) parts of the NMR signals from
spin $A$ (a, d) and spin $B$ (c, f) are shown as mesh and contour
plots as a function of $\theta$ and $\phi$; the theoretical
results are shown in (b, e).}
\end{figure*}

An important feature of the approximate cloning network is that
all input states are cloned equally well, and so it is necessary
to study the behaviour of the pulse sequence when applied to a
wide range of points on the Bloch sphere.  We have studied a total
of 312 input states, arranged on a 13 by 24 rectangular grid of
$\theta$ and $\phi$ values, with a spacing of $15^\circ$: this
choice does not cover the Bloch sphere uniformly (for example the
north and south poles are both sampled 24 times), but is
experimentally convenient.  For each input state we measured the
total NMR signals observed from spins $A$ and $B$, and their real
and imaginary components are plotted in Fig.~\ref{fig:5}.
The experimental results clearly show the expected cosine and sine
modulations, indicating that the cloning network is effective for
all these input states.

Finally, we recall that approximate quantum cloning is unitarily
equivalent to another interesting operation, the optimal \UNOT\
gate \cite{buzek00}, and so our experiment could also be
considered as an implementation of \UNOT; this point is not,
however, specifically demonstrated here.
\begin{acknowledgments}We are indebted to H.~B.~F. Dixon for suggesting the use of
ethenyl\-phos\-phonates and to M.~Bowdrey, J.~Boyd and A.~Ekert
for helpful conversations. H.K.C. thanks NSERC (Canada) and the
TMR programme (EU) for financial assistance.  J.A.J. is a Royal
Society University Research Fellow. This work is a contribution
from the Oxford Centre for Molecular Sciences, which is supported
by the UK EPSRC, BBSRC, and MRC.
\end{acknowledgments}

\end{document}